# Direct Observation of Site-specific Valence Electronic Structure at Interface: SiO$_2$/Si Interface

Y. Yamashita,[1] S. Yamamoto,[1] K. Mukai,[1] J. Yoshinobu,[1] Y. Harada,[2] T. Tokushima,[2]
T. Takeuchi,[2] Y. Takata,[2] S. Shin,[1,2] K. Akagi,[3] and S. Tsuneyuki[3]

[1]*The Institute for Solid State Physics, The University of Tokyo,*

*Kashiwa, Chiba 277-8581, Japan*

[2]*Riken/SPring-8*

*Sayo-gun, Hyogo 679-5148, Japan*

[3]*Department of Physics, Graduate School of Science, The University of Tokyo,*

*Bunkyo-ku, Tokyo 113-0033, Japan*

**Abstract**

Atom specific valence electronic structures at interface are elucidated successfully using soft x-ray absorption and emission spectroscopy. In order to demonstrate the versatility of this method, we investigated SiO$_2$/Si interface as a prototype and directly observed valence electronic states projected at the particular atoms of the SiO$_2$/Si interface; local electronic structure strongly depends on the chemical states of each atom. In addition we compared the experimental results with first-principle calculations, which quantitatively revealed the interfacial properties in atomic-scale.

Interfaces change atomic structures and chemical compositions of matters, providing not only fascinating physical properties such as metal-insulator transition [1], band gap narrowing [2], and superconductivity [3] but also affecting electronic properties of semiconductor devices [4]. Therefore, observing valence and/or conduction electronic states projected at an individual atom of the interface is in particular important to obtain atom-based picture of physical properties of matters. In spite of many studies on interfaces, the interfacial electronic states have been evaluated mostly as the average and not as individual states. Thus, what we need is a method that allows us to probe atom- specific electronic states at the interfaces directly.

Soft X-ray absorption (SXA) spectroscopy [5] is a method to study an excitation from a core level to conduction states, providing an element-specific conduction states. Soft X-ray emission (SXE) spectroscopy [6,7] probes n photon emission process involving a core hole decay process predominantly from valence states to a core level state, which addresses valence states of a particular element. Moreover, since the core electrons are localized to a particular atom we can in an atom-specific way study the valence and conduction states using SXA and SXE spectroscopy. Thus, the selective photo-absorption at interfaces is realized by tuning the incident photon energy to only allow the electronic excitation from a core level to conduction band levels, using SXA. Once the core level at the interfaces is excited, the core hole is filled by an electron from an occupied valence level accompanied with soft X-ray emission. We thus have a tool to look into nature of atom-specific electronic states at the interface directly.

In this letter, we report direct observation of the valence electronic structure projected at a particular atom of the $SiO_2/Si$ interface using SXA and SXE spectroscopy. The local electronic state are noticeably different from those of bulk $SiO_2$ and are strongly dominated by the chemical states of each atom at the interface. Furthermore, we have compared the experimental results with ab initio density functional calculations, which quantitatively reveals not only the Si-O-Si bond angle and Si-O bond length at the interface but interfacial electronic properties in atomic-scale.

$SiO_2/Si(111)$ samples were prepared from phosphorus-doped n-type Si(111) wafers that have a resistivity of 0.001 Ωcm. After standard RCA cleaning, a native oxide layer was etched away by a 1% HF solution and then the wafers were immersed in a 40% $NH_4F$ solution for 15 min to form an atomically smooth silicon surface [Si(111)(1x1)-H]. A 1.8 nm oxide layer was prepared in 0.1 M Pa of oxygen at 600 K for 5 min. It is noteworthy that the prepared oxide layer has an atomically smooth interface [8]. The synchrotron radiation experiments were performed using BL-27SU at SPring8 with the approval of JASRI as Nanotechnology Support Project (Proposal No. 2003B0209-Nsa-np-Na and 2004A0345-Nsa-np-Na). The oxygen K-edge absorption spectra were measured by detecting O KVV Auger electrons at 510 eV and an instrumental energy resolution of 50 meV. SXE spectroscopy was performed with an 800 meV energy resolution. The design and performance of the SXE spectrometer is described in detail elsewhere [9]. First-principles calculations based on the density functional theory with a generalized gradient approximated correction were also performed to evaluate the local density of states of the valence level and the interface structure and properties [10]. A six Si-layer thick slab model was used with a periodic boundary condition: the unit cell contains four dangling bonds of a clean Si(111) surface in its initial configuration. The oxidized structures were then obtained by inserting O atoms one by one with structure relaxation. We prepared three models obtained by a different insertion order of oxygen atoms and adopted the most relaxed one.

Figures 1 (a) and (b) show the oxygen K-edge absorption spectra of $SiO_2/Si(111)$ structures with 1.8 nm and 8 nm thick $SiO_2$ layers, respectively. The latter is considered to represent the bulk $SiO_2$ spectrum since the mean free path of the Auger electron is ~ 1.5nm at 510eV [11], which is much shorter than the $SiO_2$ layer thickness. An absorption spectrum of the 1.8 nm thick $SiO_2/Si(111)$ structure is strikingly different from the bulk $SiO_2$ spectrum. The 1.8 nm spectrum has a lower onset with edge structures at 530 and 531.5 eV (see the inset of Fig. 1).

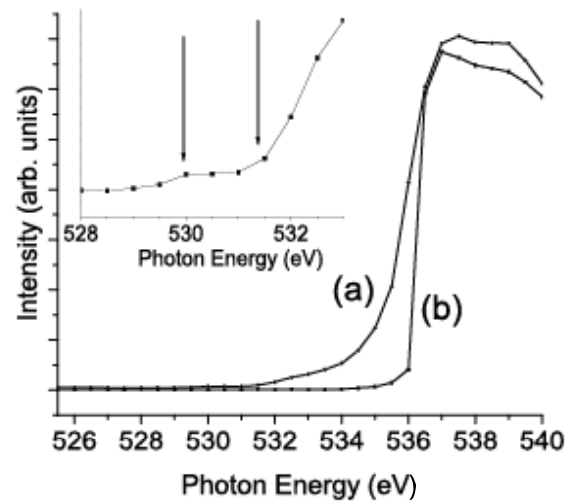

Figure 1. Oxygen K-edge absorption spectra of (a) 1.8 nm thick $SiO_2/Si(111)$ and (b) 8nm-thick $SiO_2/Si(111)$ structures. The inset shows the magnified spectrum of (a). Incident photon angle was 36° from surface normal and p-polarized light was used. 8nm-thick $SiO_2/Si(111)$ structure was prepared in 0.1 M Pa of oxygen at 1100 K for 1 hour. Note that absorption from O 1s orbital to O 2s orbital is forbidden due to selection rules. Thus, these absorption spectra represent unoccupied O 2p states.

Muller et al. used EELS and reported that the oxygen K-edge was lowered by 3 eV at the interface compared to the bulk $SiO_2$ [12]. Thus, the edge structures in the spectrum are attributed to unoccupied O 2p states at the interface, and the lowered edge exhibits a reduced band gap at the interface. Note that the peak at 537.5 eV is due to unoccupied O 2p states that are hybridized with Si 3s and 3p states of bulk $SiO_2$ [13,14].

As for the interface structure of $SiO_2/Si$, it is well known that the interface consists of intermediate oxidation states (namely suboxide), i.e., $Si^{1+}$ ($Si_2O$), $Si^{2+}$ (SiO), and $Si^{3+}$ ($Si_2O_3$) [8,15,16]. Analysis of the Si 2p photoelectron spectra (PES) for the 1.8 nm-thick $SiO_2/Si(111)$ indicates that the interface predominantly consists of $Si^{1+}$ and $Si^{3+}$ and the relative intensity of $Si^{2+}$ is only 3%. This is consistent with the previous results [8]. According to the theoretical studies by Wallis et al., the oxygen K-edge of the intermediate states in the amorphous silicon shifts to lower energy as the oxidation number of the adjacent silicon atoms decreases [17]. Thus, we assigned the absorption edges at 530 and 531.5 eV to an O atom bonding to $Si^{1+}$ and $Si^{3+}$ at the interface, respectively (hereafter, we denote O atom bonding to $Si^{1+}$ and $Si^{3+}$ at the interface as P1 and P3, respectively). Therefore, a site-specific SXE spectra of particular oxygen atoms at the interface could be selectively obtained using different

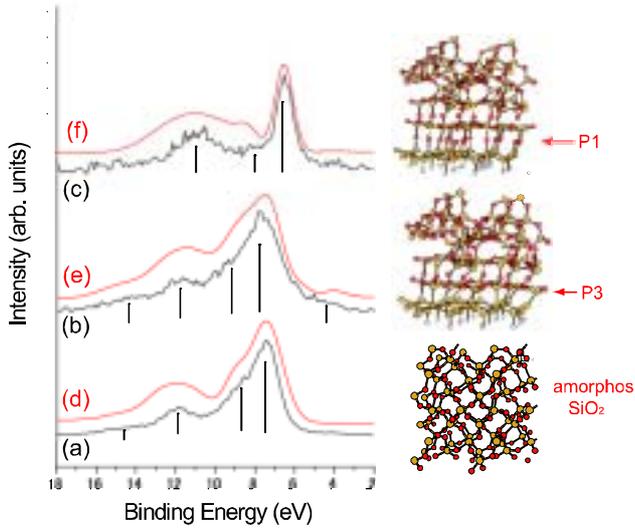

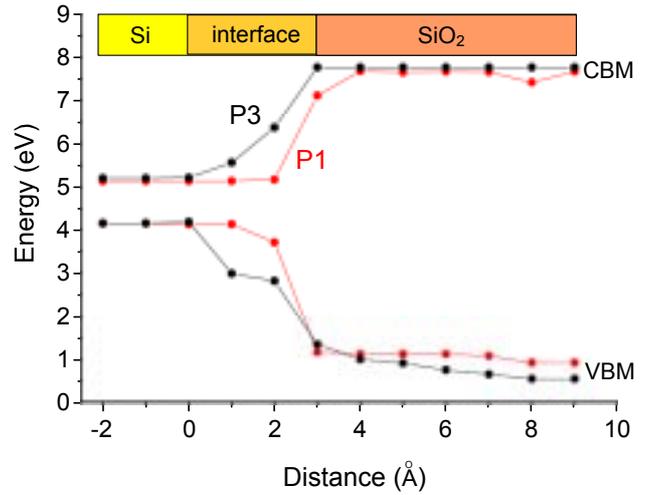

Figure 2. O K-edge SXE spectra for the 1.8 nm thick $SiO_2/Si(111)$ structure and the calculated O-2p density of states (DOS) obtained from the model presented in this figure. For the SXE spectra shown as the black lines, incident photon energies were (a) 537.5 eV, (b) 533 eV, and (c) 530 eV. Incident photon angle was 60° from surface normal and a p-polarized light was used. Note that in order to selectively excite P3, we used 533 eV as the incident light energy. Red lines show the calculated O-2p DOS for (d) $SiO_2$, (e) P3 atoms of $SiO_2/Si(111)$ interface with P3, and (f) P1 atoms of $SiO_2/Si(111)$ interface with P1. The calculated DOS spectrum was the average of each oxygen atom spectrum. Gaussian broadening was applied for the O-2p DOS using experimental and lifetime broadening (total 1.1 eV). Red balls and yellow balls correspond to oxygen and silicon atoms, respectively.

Figure 3. Calculated energy band gap variation in the structures with P1 and P3 along the normal direction to the interface. The position of the topmost silicon of the substrate, shown in Fig. 2, is the set to be the origin of the distance.

excitation energies, i.e., 530 eV for P1 and 531.5 eV for P3.

Figures 2 (a) - (c) show the O K-edge SXE spectra for the 1.8 nm thick $SiO_2/Si(111)$ structure. The electronic states at the interface are noticeably different from those of bulk $SiO_2$ and the interfacial electronic states strongly depend on the intermediate oxidation states at the interface. With a photon energy of 537.5 eV, the occupied O 2p states of bulk $SiO_2$ predominantly contribute to the SXE spectrum (Fig. 2(a)). In this spectrum, the peak at 7.25 eV and the shoulder at 8.65 eV are attributed to the O 2p non-bonding states, while the peaks at 11.75 and 14.33 eV are due to the bonding states between O 2p and Si $sp^3$ orbitals [18]. When the incident photon energy is 533 eV in order to excite P3 (Fig. 2(b)), the peaks due to non-bonding states become broader and shift towards a higher binding energy compared to bulk $SiO_2$. In addition, a small peak around 4 eV is observed. By exciting P1 (Fig 2 (c)), the peak at 6.50 eV, which is attributed to the non-bonding states of O 2p, is very sharp, indicating a localized non-bonding state.

In order to quantitatively evaluate the interface states, the experimental results are directly compared with theoretical calculations, which provides atomic-scale information on the interface properties. For the $SiO_2/Si(111)$ interface, the interface is abrupt, and $Si^{1+}$ and $Si^{3+}$ species dominate [8]. Therefore, the atomically flat $SiO_2/Si(111)$ interface that only consists of $Si^{1+}$ or $Si^{3+}$ is used as the interface model for the calculations. Figures 2(d)-(f) show the oxygen projected density of states (O-2p DOS) for $SiO_2$, P3, and P1, respectively, calculated with the model structures illustrated in the right.

The SXE spectrum of the bulk oxide region far from the interface is well reproduced by the calculated O-2p DOS for an amorphous structure, which is distorted from a cristobalite while maintaining a six-membered ring network (Fig. 2(a), (d)). The Si-O-Si angle and Si-O bond length are predominantly distributed from 140° to 155° and 1.55Å to 1.65Å in the present calculations, respectively, which are consistent with previous experimental results [19]. As for the P3 interface, the Si-O-Si angle is distributed from 131.6° to 151.6° and the Si-O bond length from 1.58Å to 1.76Å in the theoretically optimized structure. The local variation from the amorphous $SiO_2$ is caused by stress, which results in the appearance of a shoulder at 4 eV and the broadening of the non-bonding state from 7.75 eV to 9.13 eV (Fig. 2(b), (e)). This implies that the P3 layer itself is about to become an amorphous state. On the other hand, for the P1 interface, the Si-O-Si angle and Si-O bond length should be artificially maintained near 180° and 1.9Å, respectively, in order to reproduce the sharp non-bonding peak at 6.50 eV (Fig. 2(c), (f)).

As described above, we obtained interface structures with different chemical environments. Therefore, we have investigated how the local interface structures exhibit the characteristics in the interfacial region. Figure 3 shows the calculated conduction band minimum (CBM) and valence band maximum (VBM) in the interfacial region. In the interfacial region from 0Å to 2Å, the P1 interface exhibits a similar band gap to the bulk Si, while the gap of the P3 interface gradually rises. This means that the P1 interface is relatively more conductive. Thus, the interface properties strongly depend on the local structures; silicon based devices where the gate width will become less than 10 nm are essentially affected by the suboxide composition at the interface.

In this letter, we present our approach using SXA and SXE, which successfully observes the atom-specific electronic states at the $SiO_2$/Si interface. In future experiments, since SXA and SXE is a photo-in and photo-out process, the measurements can be performed even under applied electric field, magnetic field, and high pressure. Thus, the present approach should be widely applicable to interfacial electronic states of various systems and will be indispensable for evaluating atomic-scale properties of interfaces of matters and designing new exotic and fascinating materials.

In summary, using $SiO_2$/Si interface as a prototype, we successfully observed site-specific valence electronic states at the interface by means of SXA and SXE spectroscopy. We found that the interfacial electronic states were noticeably different from those of bulk $SiO_2$ and were strongly dominated by the chemical states of each atom at the interface. In addition the comparison of the experimental results with first-principle calculations quantitatively revealed not only the Si-O-Si bond angle and Si-O bond length at the interface but interfacial electronic properties in atomic-scale.

We thank Prof. Hattori and Prof. Nohira for useful discussion. This work was supported by Nanotechnology Support Project of the Ministry of Education, Culture, Sports, Science and Technology.

To whom correspondence should be addressed; e-mail: yyama@issp.u-tokyo.ac.jp